\documentclass{appolb}
\usepackage{epsfig}
\usepackage{latexsym,bm,amsmath,amssymb,amsfonts}
\usepackage{epsfig,graphics,graphicx,mathrsfs}
\usepackage{slashed,cite}
\usepackage{cite}

\newcommand{\bea}{\begin{eqnarray}}
\newcommand{\eea}{\end{eqnarray}}
\newcommand{\be}{\begin{equation}}
\newcommand{\ee}{\end{equation}}

  \renewcommand{\theequation}{\thesection.\arabic{equation}}

  \newcommand{\abar}{\bar{\alpha}_s}

  \newcommand{\rmd}{{\rm d}}

 \newcommand{\beq}{\begin{eqnarray}}
  \newcommand{\eeq}{\end{eqnarray}}

\begin{document}
\title{On high energy factorization: theoretical basics and phenomenological applications%
\thanks{Presented at Cracow Epiphany Conference, 10-12 January 2011, Krakow Poland}%
}
\author{Krzysztof Kutak
\address{Instytut Fizyki Jadrowej im. Henryka Niewodniczanskiego, \\
Polskiej Akademii Nauk,Radzikowskiego 152, 31-342 Krakow, Poland}
}
\maketitle
\begin{abstract}
We overview some of theory and phenomenology aspects of high energy factorization. In the theory part we focus on basic equations of high energy factorization i.e. BFKL, CCFM, BK. In the phenomenology part we focus on forward-central jets correlations at Large Hadron Collider and  on production of charged particles in HERA.
\end{abstract}
  \section{Introduction}
Large Hadron Collider (LHC) is already operational and Quantum Chromodynamics (QCD) is the basic theory which is used to set up the initial conditions for the collisions at the LHC but also to calculate hadronic observables. 
Application of perturbative QCD relies on so called factorization theorems which allow to decompose given process into long distance part called parton density and short distance part called matrix element. Here we will focus on high energy factorization \cite{Catani:1990eg} (there exist also collinear factorization scheme but we will not discuss it here) 
which applies when both momentum scale and energy scale involved in scattering process are high (for recent works on relation of high energy factorization to other schems we refer a reader to \cite{Dominguez:2011wm}) . 
The evolution equations of high energy factorization sum up logarithms of energy accompanied by coupling constant $\alpha_s^n \ln^m s$.
Depending on the energy range and observable one may use: BFKL \cite{Kuraev:1977fs,Balitsky:1978ic,Mueller:1993rr}, BK \cite{Balitsky:1995ub,Kovchegov:1999yj} or CCFM \cite{Ciafaloni:1987ur,Catani:1989sg,Catani:1989yc} evolution equation. When the energies of the collision are of the order of $10^3$ GeV and one considers inclusive processes in electron-proton Deep Inelastic Scattering as for example at HERA the BFKL approximation applies. 
CCFM equation since it depends on the hardness of the probe allows additionally for studies of exclusive final states.
However, if one would like to account for formation of dense system like in nuclei-nuclei collision where partons eventually overlap the BK equation or some nonlinear extension of CCFM has to be considered since it apart from splittings of gluons allows for their recombination.\\
The paper is organized as follows. In the following section 2 we introduce framework of high energy factorization, we introduce basic evolution equations. In section 3 we present phenomenological applications on two examples: production of forward-central jets at Large Hadron Collider and production of charged particles at HERA.
\section{High energy factorization and evolution equations in pQCD}
\begin{figure}[t!]
  \begin{picture}(490,100)
    \put(75, -50){
      \includegraphics{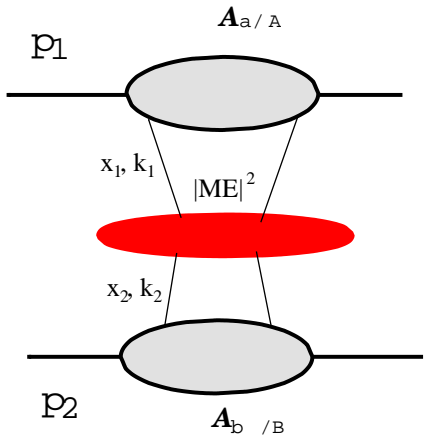}
    }
 \end{picture}
\vspace{1.3cm}
\caption{\em Factorized structure of the cross section.}
\label{fig:crossection1}
\end{figure}
The high energy factorization formula while applied to jet production in hadron-hadron scattering reads:
\bea
  \frac{d\sigma}{dy_1dy_2d^2p_{1t}d^2p_{2t}} 
  & = &
  \sum_{a,b,c,d}\int\frac{d^2k_1}{\pi}\frac{d^2k_2}{\pi}
  \frac{1}{16\pi^2 (x_1x_2 S)^2}
  \overline{|{\cal M}_{ab\to cd}|}^2\\ 
  & &
  \hskip 30pt
  \times\ \delta^2(\vec k_1 + \vec k_2- \vec p_{3t}-\vec p_{4t}){\cal A}_{a/A}(x_1,k_1^2,\mu^2)\, {\cal A}_{b/B}(x_2,k_2^2,\mu^2)\,\nonumber\\
 & &
  \hskip 30pt \times\frac{1}{1+\delta_{cd}}\nonumber,
\eea
where $k_1\equiv|{\bf k_1}|$, $k_2\equiv|{\bf k_2}|$ and $x_1$, $x_2$ are longitudinal momentum fractions see Fig(\ref{fig:crossection1}). The functions ${\cal A}_{a/A}(x_1,k_1^2,\mu^2)$ and ${\cal A}_{b/B}(x_2,k_2^2,\mu^2)$ are the unintegrated distributions which are solutions of high energy factorisable evolution equations.
They describe distributions of transversal and longitudinal momenta of partons in the incoming protons $A$ and $B$ respectively. This factorization formula apart from summing up logarithms of energy accounts also for hardness of the process. The sum is made over all flavors of initial and final partons. Similar formula can be written for DIS process.
\subsection{BFKL equation}
The simplest of high energy factorizable evolution equations is the BFKL equation. At leading order in $\ln1/x$ (LLln1/x) it reads:
\be
\frac{\partial \phi(x,k^2)}{\partial\ln1/x}=\overline\alpha_s\int_0^\infty dl^{ 2}\left[\frac{l^{2}\phi(x,l^{2})-k^2\phi(x,k^2)}{|l^{2}-k^2|}+\frac{k^2\phi(x,k^2)}{\sqrt{4l^{2}+k^2}}\right]
\ee
where we used standard notation $\phi(x,k^2)$ for factorization scale independent unintegrated gluon density. 
The real emission part of the kernel describes radiation of gluons which are strongly ordered in longitudinal momentum fraction
i.e. they are well separated in rapidity. The virtual part removes singularity when $k=l$ and is called the Regge trajectory of the gluon. This equation predicts strong rise of gluon density at small $x$ $\phi(x,k^2)=x^{-\lambda}$ and this tendency is not changed even if subleading logarithms in $\ln1/x$ are taken into account \cite{Fadin:2005zj}. Recently the BFKL
equation with special ansatz for running coupling constant and renormalization group improved kernel has been used to describe $F_2$ data and very good description of proton structure function has been achieved \cite{Kowalski:2010ue}.
\subsection{CCFM equation}
The CCFM is an equation which sums up cascades of gluons under the assumption that subsequent emissions are ordered in an emission angle.
It assumes the following form:
\begin{multline}
\mathcal{A}(x, k^2, p) \; = \; \abar \int_x^1 dz
\int \frac{\rmd ^2\bar{q}}{\pi \bar{q}^2} \,
\theta (p - z\bar{q}) \, \Delta_s(p,z\bar{q})
\left ( \frac{\Delta_{ns}(z,k, q)}{z} + \frac{1}{1-z} \right )\\
\times\mathcal{A}\left(\frac{x}{z}, k', \bar{q}\right) \; .
\label{eq:ccfminteq1}
\end{multline}
The momentum variable $p$ is defined via $\bar{\xi} = p^2/(x_n^2s)$, and $k' = |\pmb{k} + (1-z)\bar{\pmb{q}}|$
the momentum $\bar{q}$ is the rescaled momentum of the real gluon, and is related 
to $q$ by $\bar{q} = q/(1-z)$. 
Here $\Delta_s$ is the Sudakov form factor which regularizes the singularity of the $1/(1-z)$ pole, while 
$\Delta_{ns}$ is the so called non-Sudakov form factor and it corresponds to virtual contribution in the BFKL equation (their detailed form is not important in this note).\\
For recent theoretical works on this equation we refer reader to \cite{Avsar:2009pf,Avsar:2010ia} while for phenomenological applications we refer reader to \cite{Deak:2010gk,Kutak:2011gq}.
The CCFM equation has been derived after observation of coherence effects in emission of gluons \cite{Catani:1989yc} and it combines information from BFKL and DGLAP and reduces to each of them in appropriate limits: BFKL in the limit when $x\rightarrow 0$ and DGLAP when $x\rightarrow 1$.
Since it depends on hardness of the probe and also on $k^2$ of the incoming gluon it might be used in studies of final states.
In particular, one can study physics of production of jets in forward direction which we overview in section 3.  
\subsection{Saturation effects: BK equation}
As it has been already remarked, if one wants to study physics at largest energies available at LHC one eventually has to go beyond BFKL, CCFM. This is because these equations were derived in an
approximation of dilute partonic system where partons do not overlap or to put it differently do not recombine.
Because of this those equations cannot be safely extrapolated towards high
energies, as this is in conflict with unitarity requirements.
To account for dense partonic systems one has to introduce a mechanism which allows partons to recombine and therefore to saturate \cite{Gribov:1984tu}. 
Existing data suggest that the phenomenon of saturation occurs in nature. The seminal example is provided by the discovery of geometrical scaling in HERA data \cite{Stasto:2000er} and more recently by geometrical scaling in production of inclusive jets at the LHC data \cite{McLerran:2010ex,Praszalowicz:2011tc}. Also the recently observed ridge-like structure in p-p collision at he LHC \cite{Khachatryan:2010gv} has been described within approach including saturation \cite{Dumitru:2010iy}.
 There are various ways to approach the problem of evolution allowing for formation of dense system 
\cite{JalilianMarian:1997jx,JalilianMarian:1997gr,JalilianMarian:1997dw,JalilianMarian:1998cb,Kovner:2000pt,Weigert:2000gi,Iancu:2000hn,Ferreiro:2001qy,Avsar:2006jy}, 
here we are interested in the one which can be directly formulated within high energy 
factorization approach \cite{Catani:1990eg}. 
In this approach one can formulate momentum space version \cite{Bartels:2007dm} of the  Balitsky-Kovchegov equation 
which sums up 
large part of important terms for saturation and which is a nonlinear extension of the BFKL equation.
The equation reads:
\be
\begin{split}
\frac{\partial \phi(x,k^2)}{\partial \ln 1/x}= \frac{N_c\alpha_s}
{\pi}\int_0^{\infty}\frac{dl^2}{l^2}
\bigg[\frac{l^2\phi(x,l^2)- k^2\phi(x,k^2)}{|k^2-l^2|}+ \frac{
k^2\phi(x,k^2)}{\sqrt{(4l^4+k^4)}}\bigg]\\
-\frac{\alpha_s^2}{R^2} \Bigg\{
\bigg[\int_{k^2}^{\infty}\frac{dl^2}{l^2}\phi(x,l^2)\bigg]^2 
+\;\phi(x,k^2)\int_{k^2}^{\infty}\frac{dl^2}{l^2}\ln\left(\frac{l^2}{k^2}\right)\phi(x,l^2)
\Bigg\}.
\label{eq:faneq}
\end{split}
\ee
where $R$ is the radius of the proton.
The nonlinear term being convolution of the triple pomeron vertex \cite{Bartels:1994jj} with gluon density allows gluons to merge apart from gluons splitting. Due to the interplay between splitting and merging of gluons the equation above generates dynamically scale which is called saturation scale $Q_s$ (equation \ref{eq:faneq} has been also obtained by transform of coordinate version of BK equation to momentum space as has been done in \cite{Kutak:2003bd}). This scale acts effectively as a mass of gluons and therefore regulates bad infrared behavior of gluon density \cite{Kutak:2011rb}. It also selects the most probable $k$ of gluon to be of order of the saturation scale. It follows from the fact that at $k=Q_{s}$ the gluon density has a maximum:
\be
Q_{s}\equiv\partial_{\ln k^2} \phi(x,k^2)=0.
\ee 

\begin{figure}[t!]
  \begin{picture}(490,100)
    \put(20, -45){
      \includegraphics{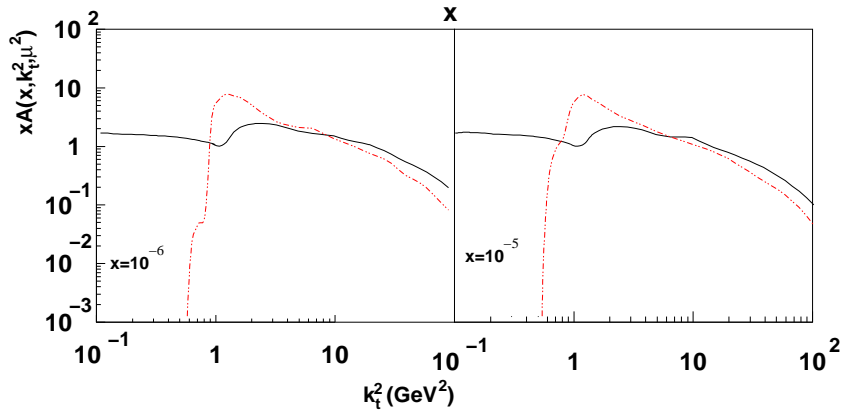}
    }
 \end{picture}
\vspace{1.3cm}
\caption{\em Gluon density obtained 
from CCFM with saturation to gluon density from CCFM as a function of $k^2$ for $x=10^{-5}$, $x=10^{-6}$ }
\label{fig:F2sat}
\end{figure}
In order to have flexible approach and be able to simulate the scattering process in detail one uses Monte Carlo implementation of evolution equations, this is the case for DGLAP, CCFM, BFKL. As BK is a nonlinear equation  it is not of straightforward usage in Monte Carlo generators. However, one can avoid complications coming from nonlinearity by applying absorptive boundary conditions \cite{Mueller:2002zm} which mimics the nonlinear term in the BK equation.
\subsection{Saturation effects: CCFM equation with absorptive boundary}
The basic principle of this method is that the absorptive boundary  limits the phase space for gluons and therefore effectively acts as nonlinear term in the BK equation. In the original formulation it was
required that the BFKL amplitude  should be equal to
unity for a certain combination of $k^2$ and $x$. In discussed here approach the energy dependent cutoff on transverse gluon momenta has been imposed. It acts as 
absorptive boundary and slows down the rate of growth of the gluon density. 
In order to have description of exclusive processes and account for saturation effects one can use CCFM evolution equation together with 
absorptive boundary \cite{Mueller:2002zm}. Its certain variation has been implemented in CASCADE Monte Carlo event generator \cite{Jung:2010si}(for adaptation to CCFM of method developed in \cite{Mueller:2002zm} we refer the reader to \cite{Avsar:2009pf,Avsar:2010ia}).
In the approach of \cite{Kutak:2008ed} as the absorptive boundary the GBW \cite{GolecBiernat:1998js} critical line has been used. The condition for saturation was provided by the GBW saturation scale   
$Q_{s}=k_0(x_0/x)^{\lambda/2}$ i.e. density of gluons with momenta generated at given $x$ with transversal momenta which satisfied 
condition $k\!\!<\!\!Q_s$ was set to go to zero as $k^2$ what is in agreement with numerical solution of BK. This prescription gives gluon density
which has a maximum as a function of $k$ in agreement with results obtained from BK. However, one should add that this method is quite simplistic and within this approach one can not find the effect of saturation of saturation scale itself which has been found in \cite{Avsar:2010ia}.
\begin{figure}[t!]
  \begin{picture}(490,100)
    \put(55, -200){
      \includegraphics{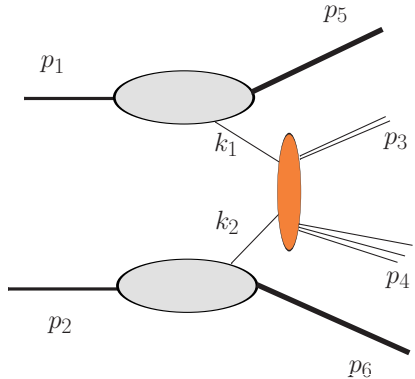}
    }
 \end{picture}
\vspace{1.3cm}
\caption{\em Production of forward-central jets.}
\label{fig:crossection}
\end{figure}
\section{Phenomenological applications.}
\subsection{Production of central-forward jets at LHC }
Physics in the forward region at hadron colliders is traditionally dominated by soft particle production. With the advent of the LHC, forward physics  phenomenology turns  into  a  largely new field~\cite{Grothe:2009nk,Jung:2009eq,d'Enterria:2009fa}   involving both soft and hard production processes, because of  the phase space opening up at  high center-of-mass energies.  
Forward jet  production   enters the LHC physics program in an essential way both for QCD studies since one can probe dense parton systems \cite{Marquet:2003dm} and for new particle searches, e.g. 
in  vector boson fusion search channels for  the Higgs boson~\cite{VazquezAcosta:2009sx,Leney:2008di}. 
The forward  production of high-p$_{\rm{T}}$ particles brings jet physics into a 
region  characterized  by  multiple energy scales and asymmetric parton kinematics. 
Here we overview results~of \cite{Deak:2010gk} where the study of forward-central jet 
correlations  of two jets has been done. 
The results of such investigations  can serve to estimate the size of  backgrounds from  QCD radiation 
between  jets   at  large  rapidity separations for  Higgs  boson searches  in vector boson fusion channels.  
High-energy factorization allows one to decompose the cross section 
for the forward-central jet production of  Fig.~\ref{fig:crossection}  into  partonic distributions  and hard-scattering   kernels,  obtained  via the  high-energy 
projectors~\cite{Catani:1990eg}  from   the amplitudes for  the process 
$p_1 +  p_2 \to p_3 + p_4 + 2$ massless partons. 
\label{sec:trv}
\begin{figure}[t!]
  \begin{picture}(30,30)
    \put(-40, -210){
      \includegraphics{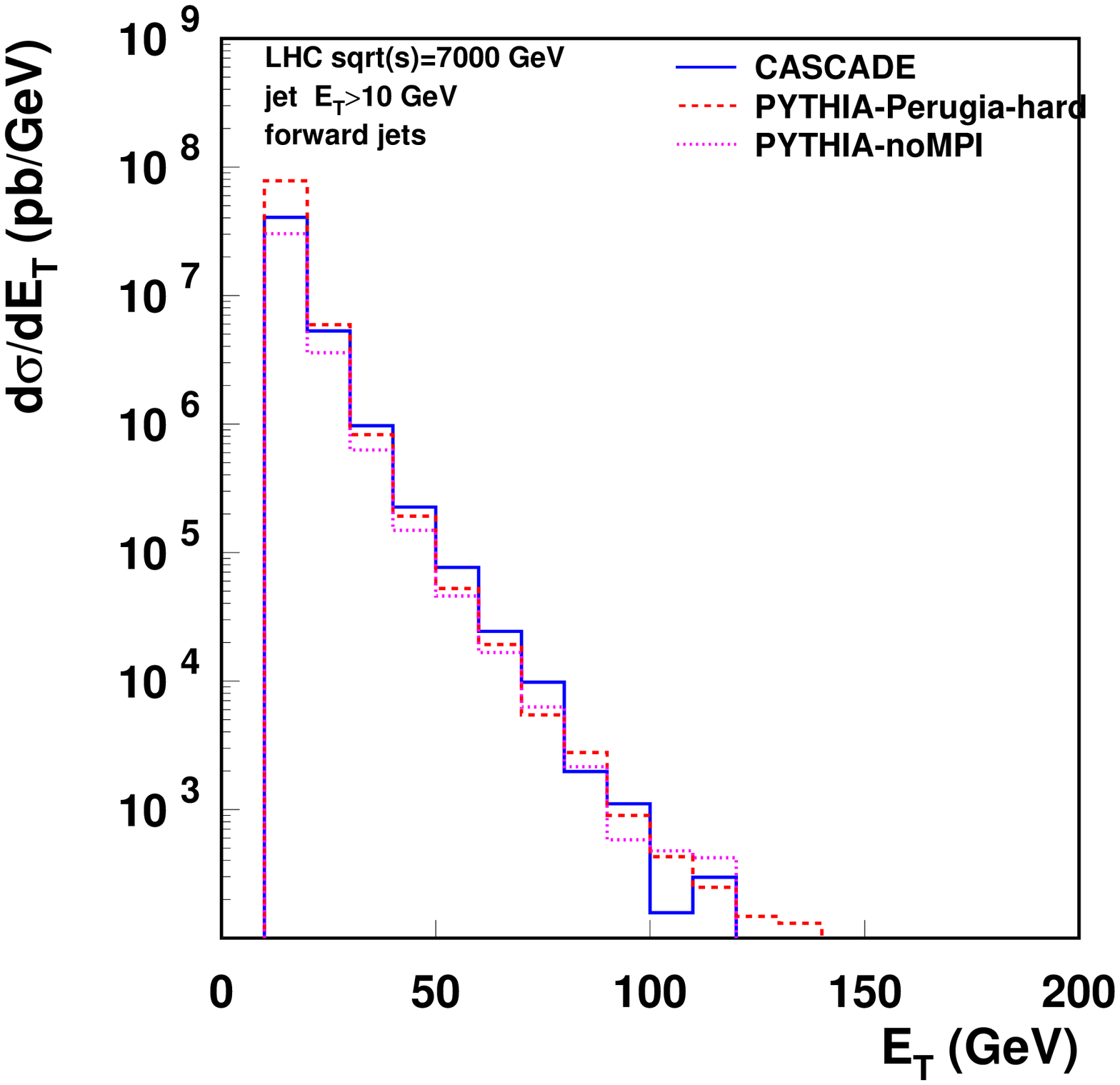}
    }
    \put(150, -210){
      \includegraphics{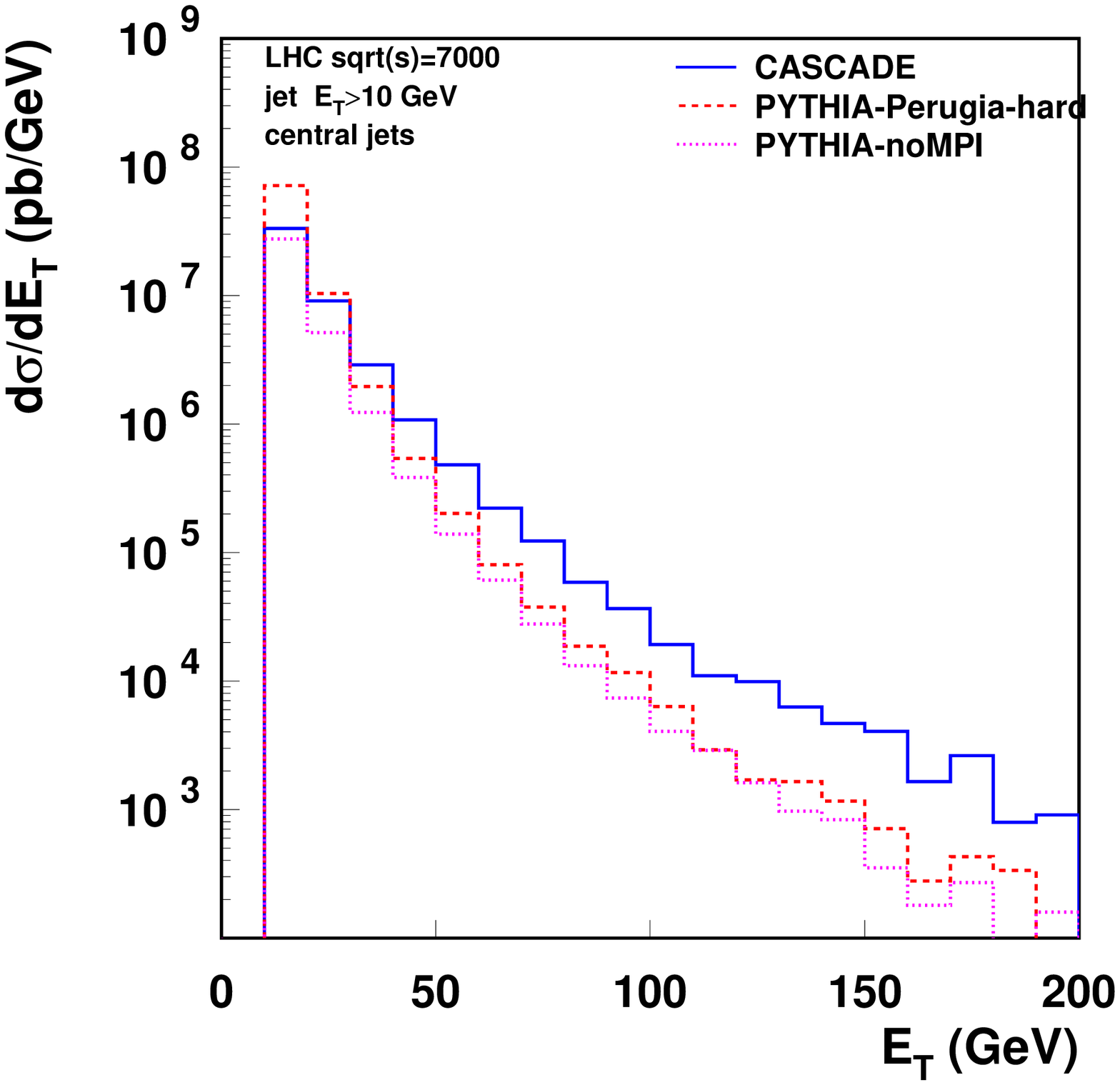}
    }

     \end{picture}
\vspace{7cm}
\caption{\em \small Transversal momentum spectra of produced jets at total collision energy $\sqrt s=7\,TeV$ with requirement that p$_\perp\!>\!10\,GeV$.
We compare predictions obtained from  CASCADE and PYTHIA running in a multiple interactions mode and no multiple interactions mode. Spectrum of forward jets (left);  
spectrum of central jets (right).}
\label{Fig:transversal}
\end{figure}
In Fig.~\ref{Fig:transversal} the prediction of differential cross section $\frac{d\sigma}{d p_{\perp}}$
is shown as obtained from CASCADE and PYTHIA. The cross sections predicted from both simulations at low momentum are of the similar order,  however, at larger transverse momentum the  CASCADE predicts a larger cross section what is clearly visible for central jets (Fig.~\ref{Fig:transversal} right).
This behavior is expected since  CASCADE uses matrix elements which are calculated within high energy factorization scheme allowing for harder transversal
momentum dependence as compared to collinear factorization. Moreover CASCADE 
applies CCFM parton shower utilizing angle dependent evolution kernel which at small $x$ does not lead to ordering in transverse momentum, 
and thus allows for more hard radiation during evolution as compared to based on leading order DGLAP splitting functions Monte Carlo generator PYTHIA. 
The parton shower has major influence on the side where the small $x$ gluon enters the hard interaction, thus the jets in the central region are 
mainly affected by the parton shower. 
\begin{figure}[t!]
\centerline{\epsfig{file=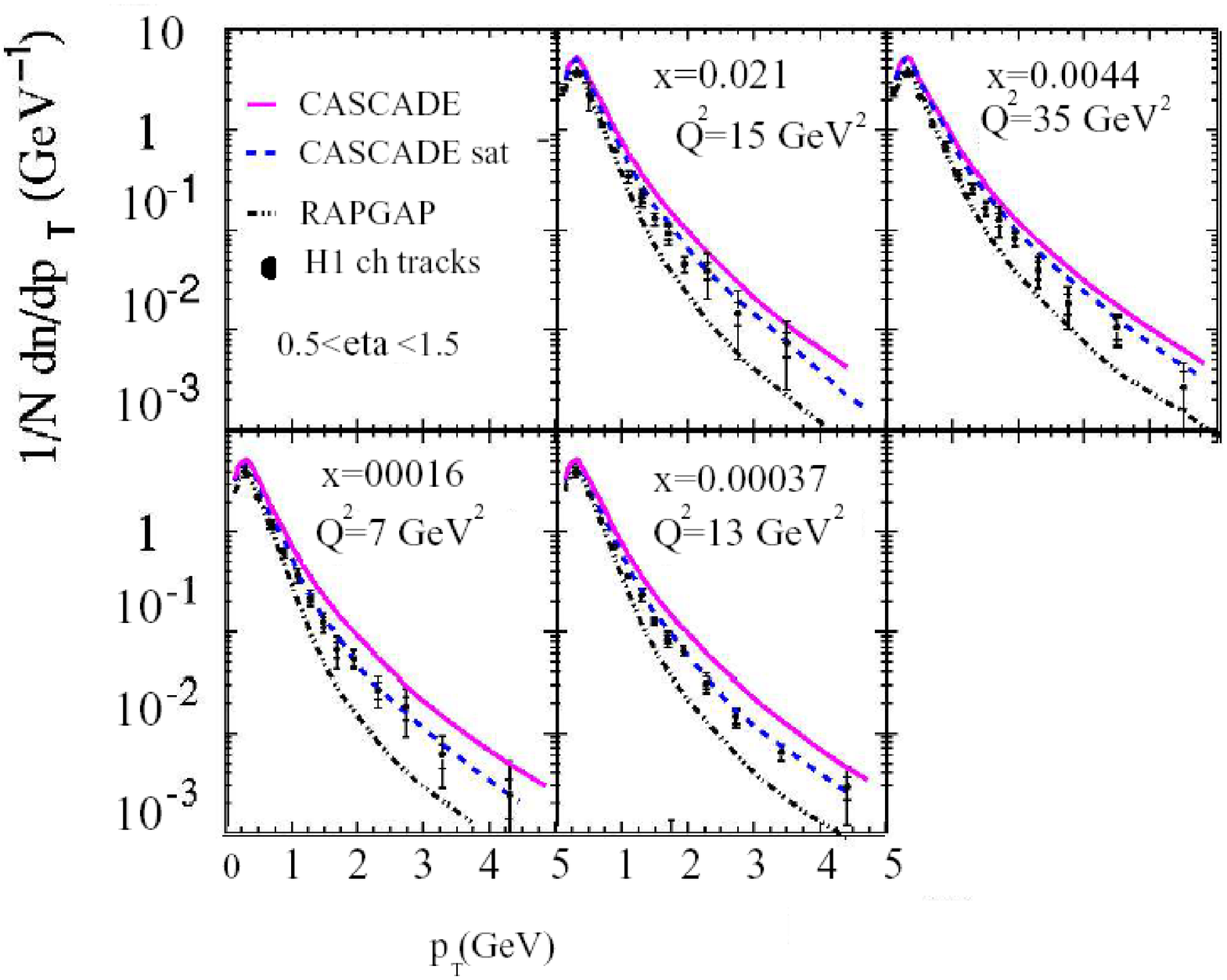,height=7cm}}
\caption{\em Differential cross section for transverse momentum distribution of charged hadrons calculated within CCFM (violet continuous line), 
CCFM with saturation (dashed blue line) and DGLAP (dotted black line)}
\vspace{-0.5cm}
\label{fig:charged}
\end{figure}
\subsection{Production of charged particles at HERA}
Another exclusive observable that is interesting to look at is the $p_T$ spectrum of produced charged particles in DIS. Here we review application of the unintegrated gluon density from CCFM with introduced saturation effects via energy dependent cut off mimicking nonlinear effects as has been done in \cite{Kutak:2008ed}. In the results we see a clear difference between the approach which includes saturation and the one which does not include
it. The description with saturation is closer to data suggesting the need for 
saturation effects \cite{Adloff:1996dy}. 
We compare our calculation with calculation based on CCFM (CASCADE) and on DGLAP (RapGAP) evolution equations. From the plots Fig. \ref{fig:charged} we see that the CCFM with saturation describes data better then the
other approaches. CCFM overestimates the cross-section for very low $x$ data
while DGLAP underestimates it. 
This is easy to explain, in CCFM one can get large contributions from larger momenta in the chain due to lack of ordering in $k^2$ while in DGLAP large $k^2$ in the chain is suppressed. On the other hand CCFM with
 saturation becomes ordered for small $x$ both in $k^2$ and rapidity and therefore
interpolates between these two. 
\section{Conclusions}
We have reviewed basic theoretical aspects of high energy factorization and gave examples of phenomenological applications to hadron-hadron collisions and lepton-proton collisions. We also stressed the uniqueness of the high energy factorization as a framework in which both hard processes and formation of dense system can be studied.
 \section*{Acknowledgements}
I would like to thank organizers of the conference Epiphany 2011 for inviting me to give a talk. Results on jets presented in this article were obtained in collaboration with M. Deak, H. Jung and F. Hautmann while results on CCFM with absorptive boundary with H. Jung. I also would like to thank S. Sapeta for useful discussions\\
This work has been supported by grant number N N202 128937


\begin{thebibliography}{99}
\bibitem{Catani:1990eg}
  S.~Catani, M.~Ciafaloni, F.~Hautmann,
  Nucl.\ Phys.\  {\bf B366 } (1991)  135-188. 


\bibitem{Dominguez:2011wm}
  F.~Dominguez, C.~Marquet, B.~W.~Xiao and F.~Yuan,
  arXiv:1101.0715 [hep-ph].


\bibitem{Kuraev:1977fs}
  E.~A.~Kuraev, L.~N.~Lipatov and V.~S.~Fadin,
  Sov.\ Phys.\ JETP {\bf 45}, 199 (1977)
  [Zh.\ Eksp.\ Teor.\ Fiz.\  {\bf 72}, 377 (1977)].

\bibitem{Balitsky:1978ic}
  I.~I.~Balitsky and L.~N.~Lipatov,
  Sov.\ J.\ Nucl.\ Phys.\  {\bf 28}, 822 (1978)
  [Yad.\ Fiz.\  {\bf 28}, 1597 (1978)].

\bibitem{Mueller:1993rr}
  A.~H.~Mueller,
  Nucl.\ Phys.\  B {\bf 415}, 373 (1994).

\bibitem{Balitsky:1995ub}
  I.~Balitsky,
  Nucl.\ Phys.\  {\bf B463 } (1996)  99-160.

\bibitem{Kovchegov:1999yj}
  Y.~V.~Kovchegov,
  Phys.\ Rev.\  D {\bf 60} (1999) 034008.


\bibitem{Avsar:2009pf}
  E.~Avsar and E.~Iancu,
  Nucl.\ Phys.\  A {\bf 829} (2009) 31.

\bibitem{Avsar:2010ia}
E.~Avsar and A.~M.~Stasto,
  JHEP {\bf 1006} (2010) 112.


\bibitem{Deak:2010gk}
  M.~Deak, F.~Hautmann, H.~Jung and K.~Kutak,
  arXiv:1012.6037 [hep-ph].
  
\bibitem{Kutak:2011gq}
  K.~Kutak,
  arXiv:1102.1334 [hep-ph].


\bibitem{Fadin:2005zj}
  V.~S.~Fadin and R.~Fiore,
  Phys.\ Rev.\  D {\bf 72} (2005) 014018.


\bibitem{Kowalski:2010ue}
  H.~Kowalski, L.~N.~Lipatov, D.~A.~Ross and G.~Watt,
  Eur.\ Phys.\ J.\  C {\bf 70} (2010) 983.

\bibitem{Ciafaloni:1987ur}
  M.~Ciafaloni,
  Nucl.\ Phys.\  B {\bf 296}, 49 (1988).

\bibitem{Catani:1989sg}
  S.~Catani, F.~Fiorani and G.~Marchesini,
  Nucl.\ Phys.\  B {\bf 336}, 18 (1990).

\bibitem{Catani:1989yc}
  S.~Catani, F.~Fiorani and G.~Marchesini,
  Phys.\ Lett.\  B {\bf 234}, 339 (1990).
 
\bibitem{Gribov:1984tu}
  L.~V.~Gribov, E.~M.~Levin and M.~G.~Ryskin,
  Phys.\ Rept.\  {\bf 100}, 1 (1983).


\bibitem{Stasto:2000er}
  A.~M.~Stasto, K.~J.~Golec-Biernat, J.~Kwiecinski,
  Phys.\ Rev.\ Lett.\  {\bf 86 } (2001)  596-599.



\bibitem{McLerran:2010ex}
  L.~McLerran and M.~Praszalowicz,
  Acta Phys.\ Polon.\  B {\bf 41} (2010) 1917.

\bibitem{Praszalowicz:2011tc}
  M.~Praszalowicz,
   ARXIV:1101.0585;

\bibitem{Khachatryan:2010gv}
  V.~Khachatryan {\it et al.} [ CMS Collaboration ],
  JHEP {\bf 1009 } (2010)  091.


\bibitem{Dumitru:2010iy}
  A.~Dumitru, K.~Dusling, F.~Gelis, J.~Jalilian-Marian, T.~Lappi, R.~Venugopalan,
  Phys.\ Lett.\  {\bf B697 } (2011)  21-25.



\bibitem{JalilianMarian:1997jx}
  J.~Jalilian-Marian, A.~Kovner, A.~Leonidov and H.~Weigert,
  Nucl.\ Phys.\  B {\bf 504}, 415 (1997).

\bibitem{JalilianMarian:1997gr}
  J.~Jalilian-Marian, A.~Kovner, A.~Leonidov and H.~Weigert,
   ``The Wilson renormalization group for low x physics: Towards the high
  Phys.\ Rev.\  D {\bf 59}, 014014 (1999).

\bibitem{JalilianMarian:1997dw}
  J.~Jalilian-Marian, A.~Kovner and H.~Weigert,
   ``The Wilson renormalization group for low x physics: Gluon evolution at
  Phys.\ Rev.\  D {\bf 59}, 014015 (1999).

\bibitem{JalilianMarian:1998cb}
  J.~Jalilian-Marian, A.~Kovner, A.~Leonidov and H.~Weigert,
   ``Unitarization of gluon distribution in the doubly logarithmic regime at
  Phys.\ Rev.\  D {\bf 59}, 034007 (1999)
  [Erratum-ibid.\  D {\bf 59}, 099903 (1999)].

\bibitem{Kovner:2000pt}
  A.~Kovner, J.~G.~Milhano and H.~Weigert,
   ``Relating different approaches to nonlinear QCD evolution at finite gluon
  Phys.\ Rev.\  D {\bf 62}, 114005 (2000).

\bibitem{Weigert:2000gi}
  H.~Weigert,
  Nucl.\ Phys.\  A {\bf 703}, 823 (2002).

\bibitem{Iancu:2000hn}
  E.~Iancu, A.~Leonidov and L.~D.~McLerran,
  Nucl.\ Phys.\  A {\bf 692}, 583 (2001).

\bibitem{Ferreiro:2001qy}
  E.~Ferreiro, E.~Iancu, A.~Leonidov and L.~McLerran,
  Nucl.\ Phys.\  A {\bf 703}, 489 (2002).


\bibitem{Avsar:2006jy}
  E.~Avsar, G.~Gustafson and L.~Lonnblad,
  JHEP {\bf 0701}, 012 (2007).
 
\bibitem{Bartels:2007dm}
  J.~Bartels, K.~Kutak,
  Eur.\ Phys.\ J.\  {\bf C53 } (2008)  533-548.

\bibitem{Bartels:1994jj}
  J.~Bartels, M.~Wusthoff,
  Z.\ Phys.\  {\bf C66 } (1995)  157-180.

\bibitem{Kutak:2003bd}
  K.~Kutak and J.~Kwiecinski,
  Eur.\ Phys.\ J.\  C {\bf 29} (2003) 521
  [arXiv:hep-ph/0303209].




\bibitem{Kutak:2011rb}
  K.~Kutak,
  arXiv:1103.3654 [hep-ph].

\bibitem{Mueller:2002zm}
  A.~H.~Mueller, D.~N.~Triantafyllopoulos,
  Nucl.\ Phys.\  {\bf B640 } (2002)  331-350..

\bibitem{Jung:2010si}
  H.~Jung, S.~Baranov, M.~Deak, A.~Grebenyuk, F.~Hautmann, M.~Hentschinski, A.~Knutsson, M.~Kramer {\it et al.},
  Eur.\ Phys.\ J.\  {\bf C70 } (2010)  1237-1249..

\bibitem{Kutak:2008ed}
  K.~Kutak, H.~Jung,
  Acta Phys.\ Polon.\  {\bf B40 } (2009)  2063-2070..


\bibitem{GolecBiernat:1998js}
  K.~J.~Golec-Biernat, M.~Wusthoff,
  Phys.\ Rev.\  {\bf D59 } (1998)  014017.


\bibitem{Grothe:2009nk}
  M.~Grothe,
  PoS {\bf 2008LHC}, 063 (2008).

\bibitem{Jung:2009eq}
  Z.~J.~Ajaltouni {\it et al.},
  arXiv:0903.3861 [hep-ph].

\bibitem{d'Enterria:2009fa}
  D.~d'Enterria,
  arXiv:0911.1273 [hep-ex].

\bibitem{Marquet:2003dm}
  C.~Marquet, R.~B.~Peschanski,
  Phys.\ Lett.\  {\bf B587 } (2004)  201-210..

\bibitem{VazquezAcosta:2009sx}
  M.~L.~Vazquez Acosta  [CMS Collaboration],
  PoS {\bf 2008LHC}, 023 (2008).

\bibitem{Leney:2008di}
  K.~J.~C.~Leney  [ATLAS Collaboration],
  arXiv:0810.3144 [hep-ex].

\bibitem{Adloff:1996dy}
  C.~Adloff {\it et al.} [ H1 Collaboration ],
  Nucl.\ Phys.\  {\bf B485 } (1997)  3-24..





\end{thebibliography}
\end{document}